\documentstyle[12pt,epsf]{article}
\begin{document}
\newfont{\blackb}{msbm10 scaled 1200}
\def\BR{\mbox{\blackb R}}
\def\BC{\mbox{\blackb C}}

\title{Noise-amplitude dependence of the invariant density for
noisy, fully chaotic one-dimensional maps} 
\author{S. Seshadri, V. Balakrishnan and S. Lakshmibala\\ 
{\em Department of Physics, Indian Institute of Technology - Madras,}\\ 
{\em Chennai 600 036, India}} 
\date{}
\maketitle 
\begin{abstract} 
We present some analytic, non-perturbative results for the invariant
density $\rho(x)$ for noisy one-dimensional maps at fully developed
chaos. Under periodic boundary conditions, the Fourier expansion
method is used to show precisely how noise makes $\rho(x)$ absolutely
continuous and smoothens it out. Simple solvable models are used to
illustrate the explicit dependence of $\rho(x)$ on the amplitude
$\eta$ of the noise distribution, all the way from the case of zero noise
($\eta \rightarrow 0$) to the completely noise-dominated limit ($\eta
=1$).
\end{abstract} 
\newpage
\section{INTRODUCTION}
\hskip .5cm One-dimensional (1D) maps exhibiting chaotic behaviour
have been used as effective models to describe a wide variety of
physical processes ranging from irregular behaviour in electronic
circuits and chemical reactions to turbulence [1,2]. In particular,
probabilistic or statistical approaches (involving, for instance,
invariant measures on attractors) enable us to by-pass the limitations
imposed by extreme sensitivity to initial conditions, and compute
various quantities of interest in terms of statistical averages [3-5].
A considerable body of mathematically rigorous results on 1D discrete-time
dynamics is also available [6,7].
As random noise is inevitably present in physical systems, much effort
has been put in to understand the effects of noise upon different
aspects of chaotic dynamics, including invariant densities and related
quantities [5, 8-17]. As much (but not all) of this work is based on
numerical analysis or simulation, analytical results are of
importance - especially so if they are non-perturbative in nature, i.e.,
if they are valid for arbitrarily large noise amplitudes.

In this paper, we re-visit 1D maps in the regime of fully-developed
chaos, in the presence of uncorrelated noise [18,19]. With the addition of
noise,
the evolution equation becomes a stochastic difference equation,
thereby making the system effectively infinite-dimensional. Imposing
periodic boundary conditions, the Fourier transform method [10,11] is
used to establish some non-perturbative results for the invariant
density $\rho(x)$. In Sec. 2, we show how the addition of noise of
arbitrarily small amplitude can make $\rho(x)$ absolutely continuous in
the interval. In Sec. 3, we consider a noise density that has a
single non-trivial Fourier mode, and show that $\rho(x)$ is also
locked in at the same mode, but with a phase shift. In Sec. 4 we
examine the dependence of $\rho(x)$ on the amplitude of the noise
density, varying the latter over the full range from zero (the 
noise-free limit) to unity (the completely noise-dominated limit). We
conclude with a few remarks on the effect of noise on the Lyapunov
exponent.
  
\section{CONTINUITY OF THE INVARIANT DENSITY}

We consider 1D endomorphisms $x_{n+1}=f(x_n)\,,\; n=0,1,\cdots\,,\; x_n
\in [-1,1]\, \equiv\, \mbox{I}$, exhibiting fully developed chaos with
invariant density $\rho^{(0)} (x)$. In the presence of additive noise,
\begin{eqnarray}
x_{n+1} &=& f(x_n)\,+\, \xi_{n+1}
\end{eqnarray}
where $\langle \xi_n \rangle \, = \, 0,\; \langle \xi_n \, \xi_{n'} 
\rangle \,=\, \langle \xi^2 \rangle \, \delta_{nn'}$. We use periodic
boundary conditions, so that the normalized noise distribution $g(\xi)$ is
also a periodic function with fundamental interval I. The invariant
density $\rho(x)$ for the noisy map (1) satisfies the {\em perturbed}
Frobenius - Perron equation [5]
\begin{eqnarray}
\rho(x) &=& \int_{\mbox{I}} \;dy\, g(x-f(y)) \, \rho(y)\;.
\end{eqnarray}
The Fourier expansions of $\rho$ and $g$ are 
\begin{equation}
\rho(x) = \frac{1}{2} \, \sum_{n=-\infty}^{\infty}\,\tilde{\rho}_n \,
\exp(i \pi nx)\;,\;g(\xi)=\frac{1}{2} \,\sum_{n=-\infty}^{\infty}\,
\tilde{g}_n \, \exp(i \pi n\xi)\,,
\end{equation}
where 
\begin{equation}
\tilde{\rho}_n\, =\, \int_{\mbox{I}} \;dx\, \rho(x)\, \exp(-i \pi
nx)\;, \;  \tilde{g}_n\, =\, \int_{\mbox{I}} \;d\xi \, g(\xi)\, \exp(-i
\pi n \xi)\,.
\end{equation}
On substituting Eqs. (3) and Eqs. (4) and using the normalization of
$\rho$
and $g$ ($\Rightarrow \; \tilde{\rho}_0 \,=\, 1,\; \tilde{g}_0 \,=\,1$),
Eq. (2) becomes equivalent to the following (infinite-dimensional) 
{\em inhomogeneous} matrix equation [10,11] for $ \tilde{\rho}_n \,, \; n\ne
0\,$:
\begin{eqnarray}
\tilde{\rho}_n &=& \tilde{g}_n \, S_{n0} \,+\, {\sum_{m}}' \,
\tilde{g}_n \, S_{nm} \, \tilde{\rho}_m
\end{eqnarray}
where ${\sum}'$ stands for a summation over all non-zero integers. The
elements of the matrix $S$ are given by
\begin{eqnarray}
S_{nm} &=& \frac{1}{2} \, \int_{\mbox{I}} \;dy \,  \exp[i\pi(my-nf(y))]
\,.
\end{eqnarray}
Iteration of Eq. (5) provides a fast-converging means of numerical
solution, once the $\tilde{g}_n $ are specified from the noise
distribution concerned. The unperturbed (or noise-free) case is recovered
on replacing $g(\xi)$ by $\delta(\xi)$, i.e., by setting $\tilde{g}_n
\,=\,1$ for all $n$.

It turns out that exact solutions for the invariant density $\rho(x)$ can
be obtained from Eq. (5) in several important cases by exploiting the
symmetries, if any, that the matrix $S$ may happen to possess [17]. For
instance, if $S_{n0}\,=\, 0$ for all $n\,\ne\,0$, both $\rho^{(0)} (x)$
and $\rho (x)$ reduce to the constant density 1/2. Again, suppose the
noise density $g(\xi)$ is symmetric about its mean value zero (a
reasonable assumption), so that $\tilde{g}_n \,=\, \tilde{g}_{-n}$. Then,
if $S_{nm}$ is even [respectively, odd] in the index $m$, while $S_{n0}$
is odd [respectively, even] in the index $n$, only the leading term in the
iterative solution of Eq. (5) survives. This leads to the exact solution 
\begin{eqnarray}
\rho(x) &=& \frac{1}{2} \, \sum _{n=-\infty}^{\infty} \, \tilde{g}_n
\, S_{n0}  \exp(-i\pi nx)\,.
\end{eqnarray}
Puting in the definitions of $\tilde{g}_n$ and $S_{n0}$, Eq. (7) can be
re-written in the form
\begin{eqnarray}
\rho(x) &=& \frac{1}{2} \, \int_{\mbox{I}} \;dy\, g(x-f(y)) \,.
\end{eqnarray}
We shall use these results in Sec. 4 to study analytically the dependence
of the invariant density on the amplitude of the noise distribution.

Here, we wish to point out a simple way of understanding  precisely how the addition of  noise leads, in
general, to a {\em smoother} invariant density [20,21]. Consider the noise-free
case: The Fourier coefficients of $\rho^{(0)} (x)$ are given by
\begin{equation}
\tilde{\rho}_n^{(0)} = S_{n0} \, + \, {\sum_m}' S_{nm} \,
\tilde{\rho}_m^{(0)} = S_{n0} \,+\, {\sum_m}' S_{nm} \, S_{m0} \,+\,
{\sum_m}'\,{\sum_l}' S_{nm} \, S_{ml} \, S_{l0}\,+\, \cdots 
\end{equation}
The asymptotic (large $n$) behaviour of $\tilde{\rho}_n^{(0)}$ is thus
controlled by $S_{nm}$. If $\tilde{\rho}_n^{(0)} \sim \, n^{-1}$, then
$\rho^{(0)} (x)$ has finite discontinuities in I -- including, possibly,
the end points $\pm 1$, as we have used periodic boundary conditions. Now
consider what happens when noise is added. If $g(\xi)$ is continuous, then
$\tilde{g}_n \sim n^{-2}$. Even if $g(\xi)$ has finite jumps in I
(including, possibly, at the end points $\pm 1$), its aymptotic behaviour
is at least $O(n^{-1})$ [22]. It follows at once from Eq. (5) that the
asymptotic behaviour of $\tilde{\rho}_n$ is improved to $\tilde{\rho}_n
\,=\, O(n^{-1-\alpha})$ where $\alpha > 0$. Consequently, $\sum \,
\tilde{\rho}_n\, \exp(i \pi nx)$ converges absolutely in I, and $\rho(x)$
becomes continuous everywhere, including the end points. (That is,
$\rho(-1)\,=\, \rho(1).$) Explicit illustrations will be given
subsequently.

\section{EXACT SOLUTION  FOR `SINGLE - MODE' \\ NOISE DENSITY}  
We show now that an interesting form of ``mode-locking'' occurs if the
noise density has a single wavelength, i.e., Fourier component. The
corresponding noise density is given by the one-parameter family of
functions 
\begin{eqnarray}
g(\xi) &=& \frac{1}{2} \, [1\,+\, (-1)^{r-1} \, \cos \pi r \xi]\, ,
\end{eqnarray}
where $r$ is an integer. Correspondingly,
\begin{eqnarray}
\tilde{g}_n &=& \delta_{n,0} \, +\, \frac{1}{2} \,  (-1)^{r-1}
(\delta_{n,r}\, +\, \delta_{n,-r})\,.
\end{eqnarray}
Hence the only nonvanishing Fourier coefficients
$\tilde{\rho}_n \;(n \ne 0)$ are $\tilde{\rho}_r$ and $\tilde{\rho}_{-r}
\,=\,
{\tilde{\rho}_r}^*$. Using the relation $S_{-n,-m} \,=\, S_{n,m}^*$, we
find from Eq. (5) the solution
\begin{eqnarray}
\tilde{\rho}_r &=& \frac{2(-1)^{r-1} \, S_{r0} \,-\, S_{r0} \, S_{rr}^*\,
+\, S_{r0}^* \, S_{-rr}^*}{4-4(-1)^{r-1} \mbox{Re}\, S_{rr}\,
+\,|S_{rr}|^2\, -
\,|S_{-rr}|^2} \,.
\end{eqnarray}
Writing $\tilde{\rho}_r\,=\, |\tilde{\rho}_r| \, \exp(i\phi)$, this leads
to the result
\begin{eqnarray}
\rho(x) &=& \frac{1}{2} \,+\, |\tilde{\rho}_r|\, \cos(\pi rx+\phi)\,.
\end{eqnarray}  
The invariant density is therefore `locked in' at the same wavelength as
the noise density, with a phase shift. The amplitude and phase shift
$\phi$ depend on the map $f$ through the matrix  elements of $S$ that
appear in Eq. (12). It is evident that the situation  corresponds to
completely noise-dominated dynamics. This is brought out explicitly in the
example considered in the next section.

\section{DEPENDENCE OF $\rho(x)$ ON THE NOISE AMPLITUDE}
We want to study the explicit dependence of the invariant density on the
amplitude of the noise density in a {\em non-perturbative} manner. For
this purpose, we consider illustrative cases in which an exact solution
for $\rho(x)$ is possible owing to symmetries present in $S_{nm}$, as
indicated in Sec. 2. The noise density is taken to have a compact support
$[-\eta,\eta]$ (i.e., an amplitude $\eta$). A convenient form for the
normalized density $g(\xi)$ that enables us to scan the entire range of
$\eta$ from 0 to 1 is 
\begin{equation}
g(\xi) \,=\, \left \{ \begin{array}{cc}
                    (2\eta)^{-1} \, (1\,+\, \cos \frac{\pi \xi}{\eta})\,,
&|\xi|\, < \,\eta \\
0, & \eta \, < |\xi| \,\le 1 \end{array} \right.
\end{equation}
where $0\,< \, \eta \,<\, 1$. Figure 1 depicts $g(\xi)$ for $\eta \,=\,
0.2$. In the limit $ \eta \rightarrow \, 0$, we have  $g(\xi) \rightarrow
\,\delta(\xi)$, or
noise-free dynamics. When  $ \eta \rightarrow \,1$, we have  the fully
noise dominated case considered in the preceding section. The Fourier
coefficients of $g(\xi)$ are given by
\begin{eqnarray}
\tilde{g}_n &=& \frac{\sin\,(n\pi \eta)}{n\pi \eta} \, \left [ 1\,-\,
\frac {n^2\,\eta^2}{n^2\,\eta^2 \,-\,1} \right ]\,.
\end{eqnarray}

To bring out all the effects of noise on $\rho(x)$ which we want to
demonstrate, let us first consider the square-root cusp map (see Fig. 2)
\begin{equation}
x_{n+1} = f(x_n) = 1\,-\, 2 \, |x_n|^{1/2}\;\;, \;\; x_n \in \mbox{I}\;.
\end{equation}
It is well known that this map provides a prototypical model of
intermittent chaos, arising from the marginal stability of the fixed point
at the left boundary, $x \,=\, -1$. However, unlike many other models of
intermittency that share the latter feature, the map (16) has the
non-singular invariant density [23]
\begin{eqnarray}
\rho^{(0)} (x) &=& \frac{1\,-\, x}{2}\;\;.
\end{eqnarray}
As $f(x)$ is symmetric, we have $S_{nm} \, =\, S_{n,-m}$. Further, 
\begin{equation}
S_{n0} \,=\, \frac{1}{2} \, \int_{\mbox{I}} \, dy \, \exp(2\pi i n
\sqrt{y})\,=\,  \frac{(-1)^n}{(i\pi n)} \,=\, - S_{-n,0}\,.
\end{equation}
Hence, as pointed out in Sec. 2, the exact solution for $\rho(x)$ is
given by Eq. (7). We get
\begin{eqnarray}
\rho(x) &=& \frac{1}{2} \,+\, \sum_{n=1}^{\infty} \frac{(-1)^n}{n\pi} \,
\tilde{g}_n \, \sin(n\pi x)\,.
\end{eqnarray}
Putting in the expression for $\tilde{g}_n$ from Eq. (15) and carrying out
the summations involved, we finally obtain the closed-form expression
\begin{equation}
\rho(x) \,=\, \left \{ \begin{array} {lc}
\frac{1}{2} \,+\, \frac{1}{2\eta} \, (1-\eta) \,(1+x) \, +\,
\frac{1}{2\pi} \, \sin \, \frac{\pi (1+x)}{\eta} \, , & -1\le x \le
-1+\eta \\
\frac{1-x}{2}\,, & -1+\eta \le x\le 1-\eta \\
\frac{1}{2} \,-\, \frac{1}{2\eta} \, (1-\eta) \,(1-x) \, +\,
\frac{1}{2\pi} \, \sin \, \frac{\pi (x-1)}{\eta} \, , & 1-\eta \le x \le
1\, .
\end{array} \right.
\end{equation}
Figure 3 shows $\rho(x)$ in the cases $\eta \, =\, 0.1, \, 0.2$ and 0.7
respectively. (Throughout this paper, relatively large values of $\eta$
have been chosen for clarity of illustration, and also to emphasize the
non-perturbative nature of the results.) The following points are
noteworthy:\\
(i) Both $\rho^{(0)} (x)$ and $\rho(x)$ are antisymmetric about the mean
value 1/2. However, $\rho^{(0)} (-1)$  = 1 while $\rho^{(0)} (1) =0$.
But $\rho (x)$ is continuous everywhere, {\em including} the end points
$\pm 1\; (\rho(\pm 1) $ =  1/2), in accordance with the general result
established in Sec. 2. This feature persists for arbitrarily small values
of $\eta$. It is known that it is rather difficult in numerical
simulations to obtain the exact result $\rho^{(0)} (x) = (1-x)/2$: a
`boundary layer' persists, especially near $x=-1$, 
in which the invariant density {\em builds up} to a value close to unity, instead of straightaway starting at that value for $x=-1$ and then decreasing linearly
 as $x$ increases. 
Our result
on the continuity of $\rho(x)$ accounts for this phenomenon: noise,
albeit in the form of round-off errors, is inevitably present in simulations.
(ii) In the opposite, noise-dominated limit $\eta \rightarrow 1$, we find
\begin{eqnarray}
\rho(x) &=& \frac{1}{2} \, - \, \frac{\sin \,(\pi x)}{2\pi}\;,
\end{eqnarray}
again in accord with the results of Sec. 3, in the special case $r=1, \,
|\tilde{\rho}_r| \,=\, 1/(2\pi)\, , \, \phi \,=\, \pi /2$.

Next, let us consider a simple case in which $\rho^{(0)} (x)$ has a finite
jump in the interior of the interval, in order to see how the noise
smoothens it out. A convenient illustration is provided by the piecewise
linear map (see
Fig. 4) 
\begin{equation}
x_{n+1}\,=\, f(x_n)\,=\, \left \{ \begin{array} {lc}
1\,-\, 5|x|\,, & |x| \, \le \frac{1}{5} \\
\frac{(1\,-\, 5|x|)}{4}\,, & \frac{1}{5} \, \le \, |x| \, \le 1\,\,.
\end{array} \right.
\end{equation}
The unperturbed invariant density is the piecewise constant function
\begin{equation}
\rho^{(0)} (x) \,=\, \left \{ \begin{array} {cc} 
\frac{4}{5} \,, & -1\le\,x\, < 0\\
\frac{1}{5} \,, & 0 < x \, \le 1
\end{array} \right.
\end{equation}
with a jump at $x=0$. In this case, too, $S_{nm} \,=\, S_{n,-m}$, and 
\begin{equation}
S_{n0} \,=\, - S_{-n0} \,=\, \frac{3i}{5n \pi} [1 \,-\, (-1)^n]\,\,.
\end{equation}
When noise (distributed as in Eq. (14) is added, $\rho(x)$ is found to be
the following piecewise analytic function, for $\eta \, < 1/2$:
\begin{equation}
\rho(x) \, =\, \left \{ \begin{array} {lc} 
 \frac{1}{2}\,+\, \frac{3}{10 \eta} (1+x) \, +\, \frac{3}{10 \pi} \, \sin
\,\frac{\pi(1+x)}{\eta}\,, & -1\le x\le -1+\eta \\
\frac{4}{5}\,, & -1+\eta \le x \le -\eta \\
\frac{1}{2}\,-\, \frac{3x}{10 \eta} \, -\, \frac{3}{10 \pi} \, \sin
\,\frac{\pi x} {\eta}\,, &  |x| \le \eta \\   
\frac{1}{5}\,, & \eta \le x \le 1-\eta \\  
 \frac{1}{2}\,- \, \frac{3}{10 \eta} (1-x) \, -\, \frac{3}{10 \pi} \, \sin
\,\frac{\pi(1-x)}{\eta}\,, & 1-\eta \le x \le 1\, . \\
\end{array} \right.
\end{equation}
$\rho (x)$ remains piecewise analytic for $\eta > 1/2$, as well. As
before, we illustrate in Fig. 5 the trends in the variation of $\rho (x)$
using large values of $\eta$ : 0.2, 0.5 and 0.7 respectively. As expected,
the discontinuities that $\rho^{(0)} (x)$ has at $x=0$ and $\pm 1$
disappear in $\rho (x)$, for arbitrarily small $\eta$. As $\eta
\rightarrow 1$, $\rho(x)$ tends to $1/2 \, -\, (3/5\pi) \sin(\pi x)$, corroborating our result obtained for `single - mode' noise.

\section{CONCLUDING REMARKS}
We have shown that additive noise significantly alters the invariant
density for fully chaotic 1D maps: arbitrarily small amounts of noise
smoothen out the density and remove any discontinuities it may have in
the absence of noise, under periodic boundary conditions. One may ask
whether the Lyapunov exponent $\lambda^{(0)}$ is also altered by
the addition of noise [24]. While this {\em is} so in general, there is no
change  in $\lambda^{(0)}$  in the examples considered in Sec. 4, for the
following reason: if the map $f(x)$ is an even function of $x$, so is $\ln
\, |f' (x)|$. If, further, both $\rho^{(0)} (x) \,-\, 1/2$ and $\rho (x)
\,-\, 1/2$ are antisymmetric in $x$, it follows that the corresponding
Lyapunov exponents are equal to each other, and are given by
\begin{equation}
\lambda^{(0)} \,=\, \lambda \,=\, \int_0^1 \, dx \, \ln |f' (x)|\,.
\end{equation}

The general conclusion is as follows: if $f(x)$ is an even function in I
and so is the noise density, then $\lambda$ remains equal to
$\lambda^{(0)}$ provided $S_{n0}$ is odd in the index $n$, i.e.,
$\int_{\mbox{I}} \, dx \, \cos \, (n\pi f(x)) $ vanishes for every integer
value of $n$.

\section*{Acknowledgments}
SS acknowledges financial support from the Council of Scientific and  
Industrial Research, India, in the form of a Senior Research Fellowship.
SL thanks the Department of Science and Technology, India, for partial
assistance under the scheme SP/S-2/E-03/96.

\newpage
\section*{Figure Captions}
\begin{enumerate}
\item Plot of $g(\xi)$ vs $\xi$ for $\eta \, =\, 0.2$
\item The square-root cusp map
\item Plot of $\rho(x)$ vs $x$ for the noisy square-root cusp map
corresponding to $\eta \, =\, 0.1,\, 0.2$ and 0.7 respectively
\item The piecewise linear map
\item Plot of $\rho(x)$ vs $x$ for the piecewise linear map corresponding
to  $\eta \, =\, 0.2,\, 0.5$ and 0.7 respectively
\end{enumerate}

\newpage
\begin{figure}
\epsffile{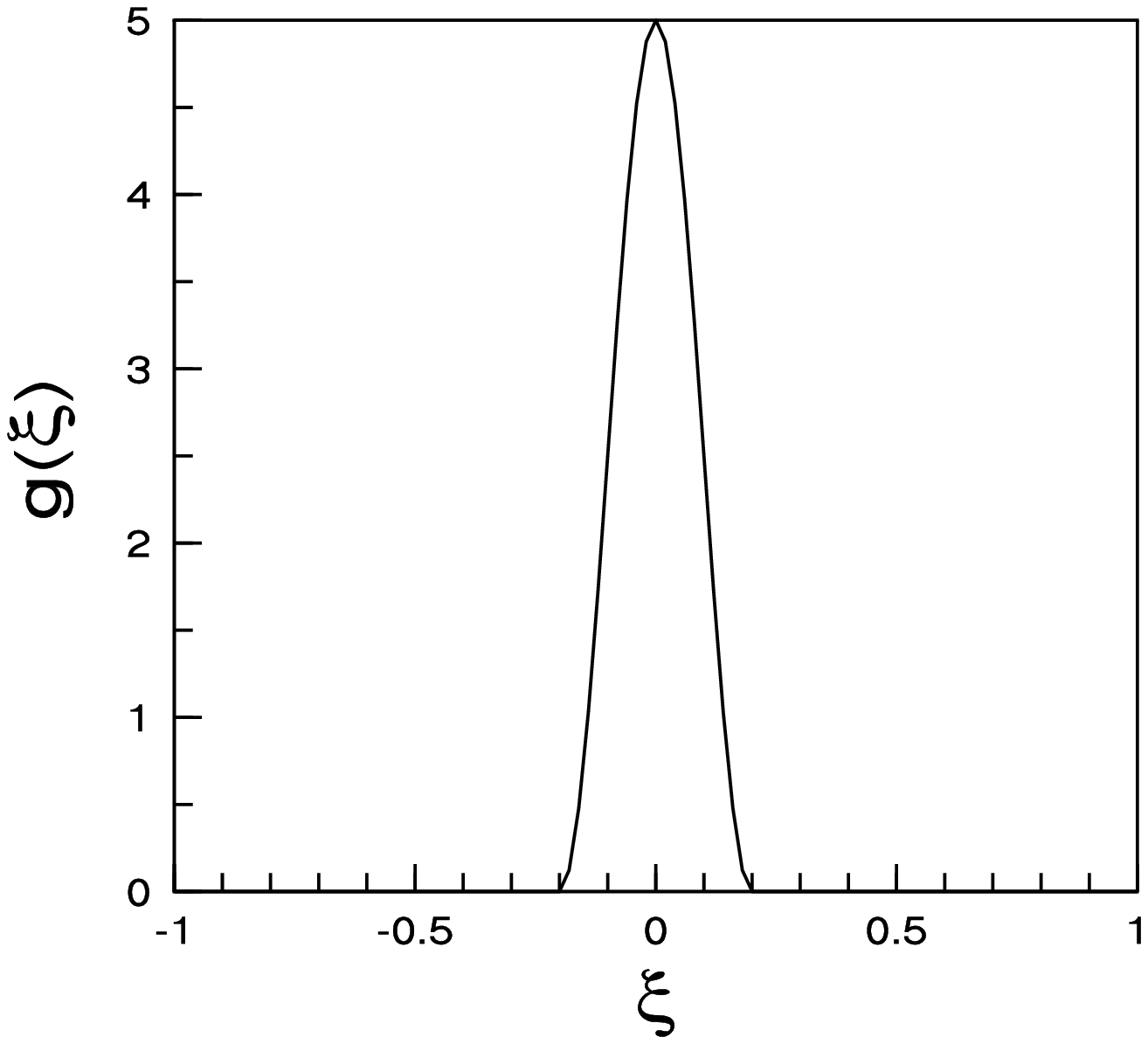}
\caption{}
\end{figure}
\newpage
\begin{figure}
\epsffile{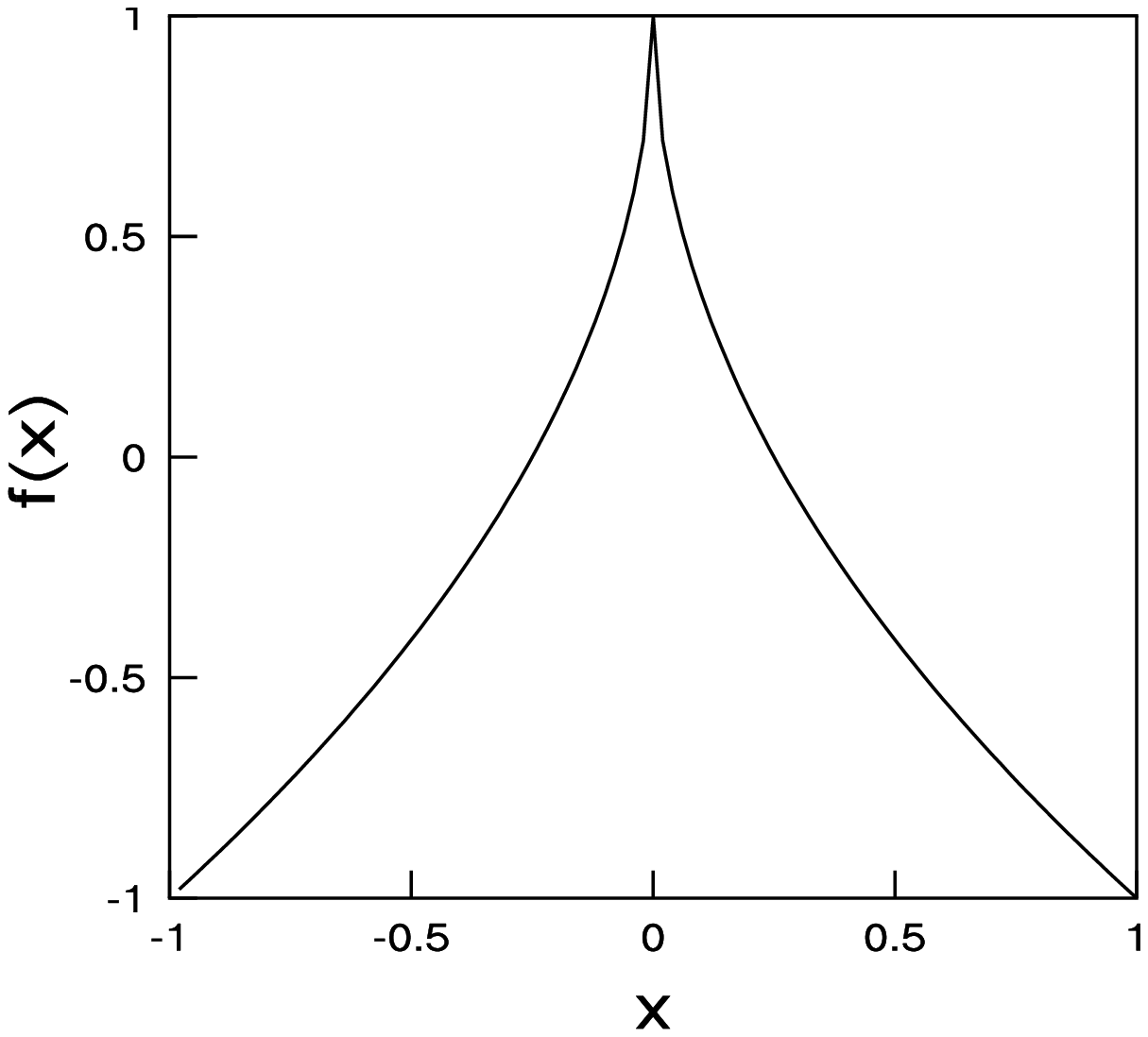}
\label{Fig. 2}
\caption{}
\end{figure}
\newpage
\begin{figure}
\epsffile{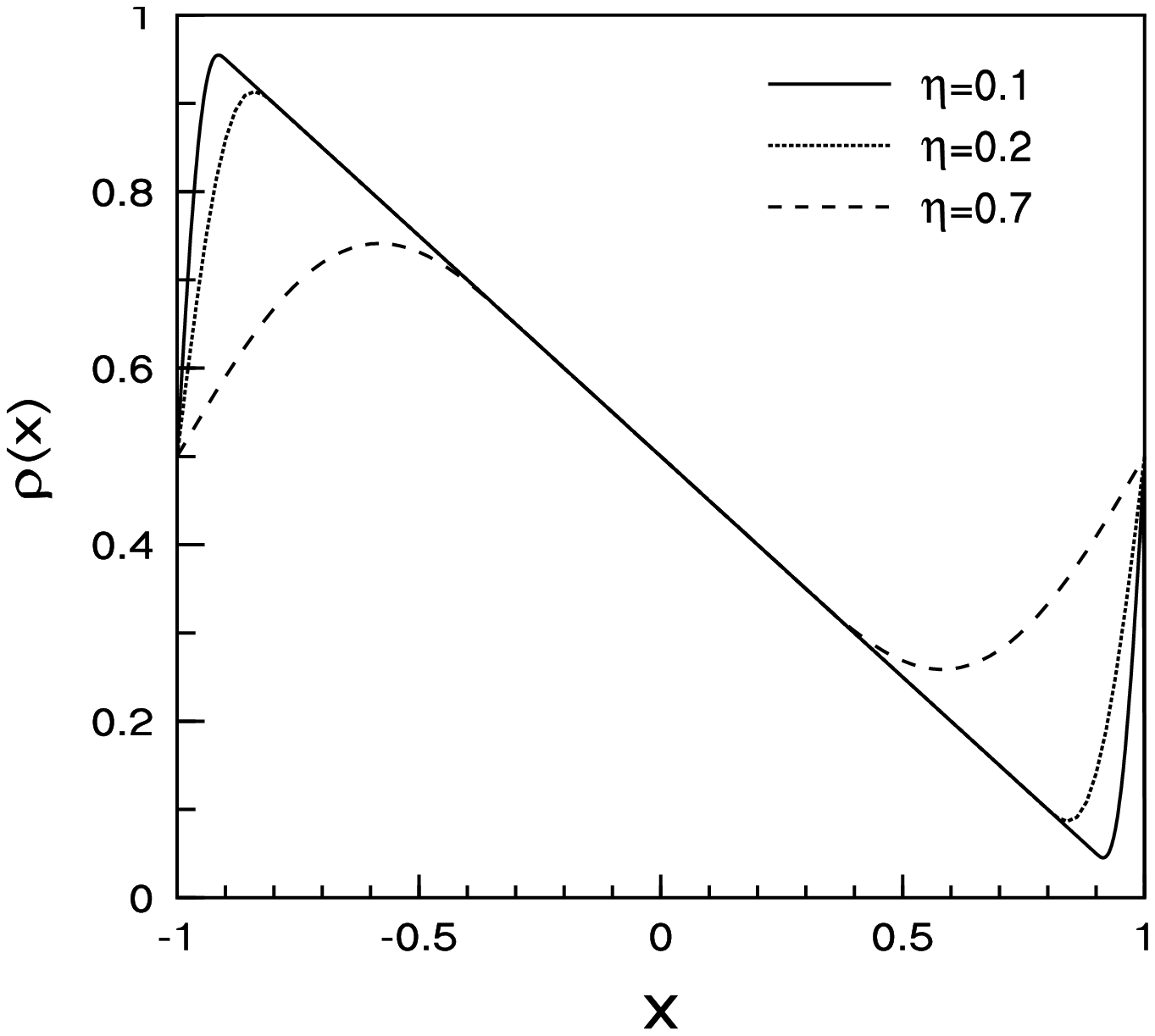}
\label{Fig. 3}
\caption{}
\end{figure}

\newpage
\begin{figure}
\epsffile{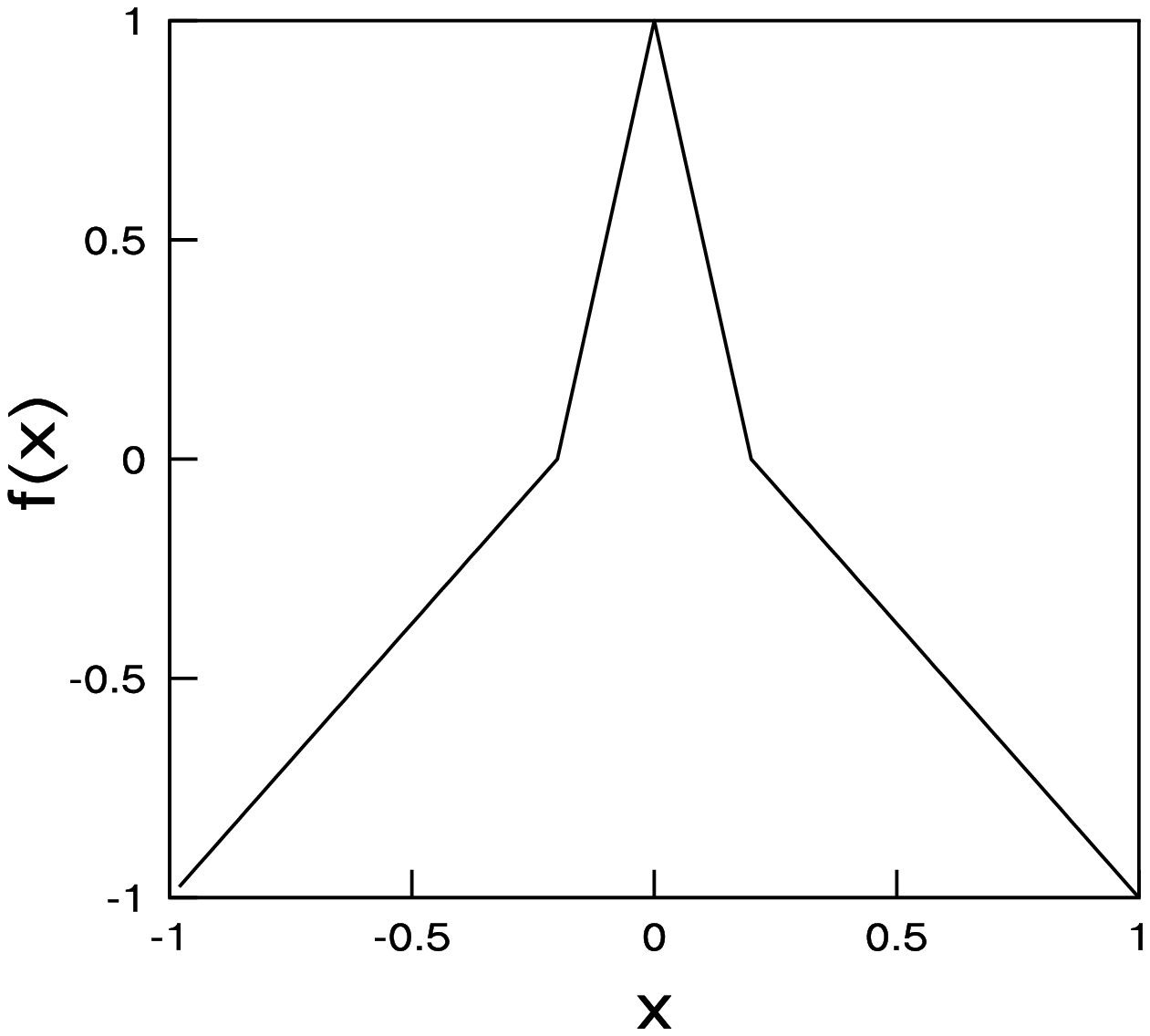}
\label{Fig. 4}
\caption{}
\end{figure}
\newpage
\begin{figure}
\epsffile{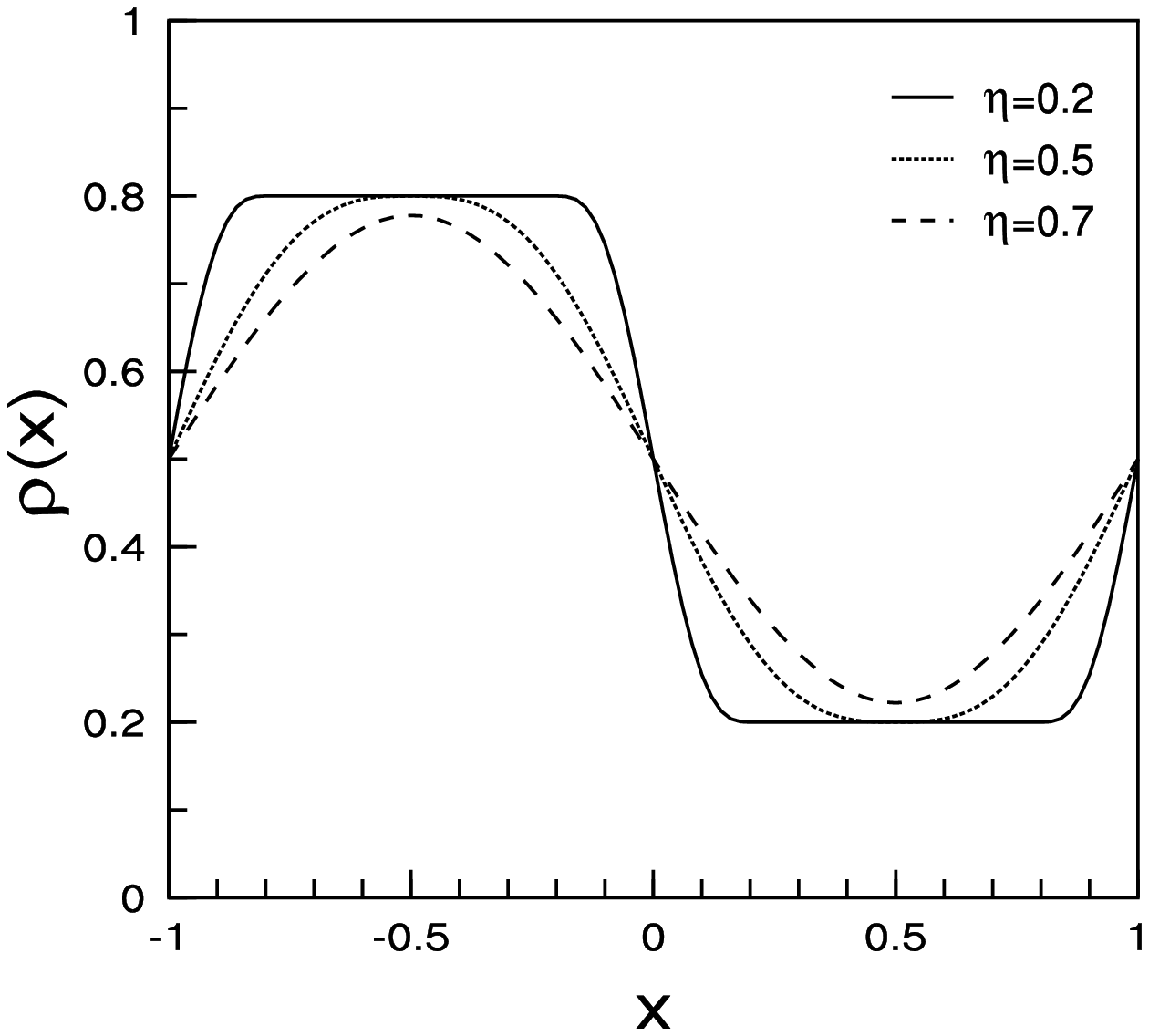}
\label{Fig. 5}
\caption{}
\end{figure}


\begin{thebibliography}{99}
\bibitem{18} If the low-dimensional dynamics arises (as it often
does) as a reduced or effective description of some `true' higher
dimensional dynamics, the inclusion of white noise in the latter will then
formally lead to colored noise in the reduced system (see , e.g., Ref.
[19]). Hence it is the effects of colored rather than uncorrelated noise
that we should consider. However, in practice this is ultimately a
question of relative time scales in the specific problem concerned: If
there is a clear separation between the characteristic time scales of the
noise and the deterministic dynamics, neglecting the correlation time of
the former may be quite justified.
\end{thebibliography}
\end{document}